\documentclass[smallextended]{svjour3}

\usepackage{url}

\usepackage{mathptmx}      
\usepackage{times}
\usepackage{graphicx}
\usepackage{listings}

\usepackage[nolists]{endfloat}
\usepackage{setspace}
\newcommand{\longversion}[1]{}
\newcommand{\shortversion}[1]{#1}

\journalname{International Journal of Parallel Programming}

\begin{document}
 
\title{Correlating Radio Astronomy Signals with Many-Core Hardware}

\author{Rob V. van Nieuwpoort \and John W. Romein}

\institute{
ASTRON, Netherlands Institute for Radio Astronomy, Oude Hoogeveensedijk 4, 7991 PD Dwingeloo, \\
The Netherlands
\email{\{nieuwpoort, romein\}@astron.nl}
}

\maketitle

\begin{abstract}
A recent development in radio astronomy is to replace traditional dishes
with many small antennas. The signals are combined to form one large,
virtual telescope.  The enormous data streams are cross-correlated to
filter out noise.  This is especially challenging, since the
computational demands grow quadratically with the number of data
streams. Moreover, the correlator is not only computationally intensive, but also
very I/O intensive. The LOFAR telescope, for instance, will produce
over 100~terabytes per day.  The future SKA telescope will even
require in the order of exaflops, and petabits/s of I/O.  A recent
trend is to correlate in software instead of dedicated hardware, to
increase flexibility and to reduce development efforts.

We evaluate the correlator algorithm on multi-core CPUs
and many-core architectures, such as NVIDIA and ATI GPUs,
and the \mbox{Cell/B.E.}  The correlator is a streaming, real-time
application, and is much more I/O intensive than applications that are
typically implemented on many-core hardware today.  We compare with
the LOFAR production correlator on an IBM Blue Gene/P supercomputer.
We investigate performance, power efficiency, and programmability.  We
identify several important architectural problems which cause
architectures to perform suboptimally.  Our findings are applicable to
data-intensive applications in general.

The processing power and memory bandwidth of
current GPUs are highly imbalanced for correlation purposes.  While
the production correlator on the Blue Gene/P achieves a superb 96\% of the
theoretical peak performance, this is only 16\% on ATI GPUs, and 32\%
on NVIDIA GPUs. The \mbox{Cell/B.E.} processor, in contrast, achieves an
excellent 92\%. We found that the \mbox{Cell/B.E.} and NVIDIA GPUs are the most
energy-efficient solutions, they run the correlator at least 4 times more energy
efficiently than the Blue Gene/P.  The research presented is an
important pathfinder for next-generation telescopes.



\end{abstract}

\section{Introduction}

A recent development in radio astronomy is to build instruments where
traditional dishes are replaced with many small and simple
omni-directional antennas. The signals of the antennas are combined to
form one large virtual telescope. Examples include current and future
instruments such as LOFAR (LOw Frequency Array)~\cite{ppopp2010},
MeerKAT (Karoo Array Telescope)~\cite{meerkat}, 
ASKAP (Australian Square Kilometre Array Pathfinder)~\cite{askap},
and SKA (Square Kilometre Array)~\cite{ska}. These new generation telescopes produce
enormous data streams. The data streams from the different antennas
must be cross-correlated to filter out noise. The correlation process
also performs a data reduction by integrating samples over time.  The
correlation step is especially challenging, since the computational
demands grow \emph{quadratically} with the number of data streams. The
correlator is extremely demanding, since it is not only
computationally intensive, but also very data intensive. In the
current field of radio astronomy, the number of operations that has to
be performed per byte of I/O is exceptionally small.
For astronomy, high-performance computing is of key
importance. Instruments like LOFAR are essentially software
telescopes, requiring massive amounts of compute power and data
transport capabilities. Future instruments, like the SKA~\cite{ska},
need in the order of exaflops of computation, and petabits/s of I/O.

Traditionally, the online processing for radio-astronomy instruments
is done on special-purpose hardware. A relatively recent
development is the use of supercomputers~\cite{spaa-06,ppopp2010}. Both
approaches have several important disadvantages. Special-purpose
hardware is expensive to design and manufacture and, equally
important, it is inflexible. Furthermore, the process from creating a
hardware design and translating that into a working implementation
takes a long time. Solutions that use a supercomputer (e.g., a Blue
Gene/P in the LOFAR case) are more flexible~\cite{ppopp2010}, but are
expensive to purchase, and have high maintenance and electrical power
cost. Moreover, supercomputers are not always well-balanced for our
needs. For instance, most supercomputers feature highly efficient
double-precision operations, while single precision is sufficient for
our applications.


\begin{figure*}[t]
\begin{center}
\includegraphics[width=12cm]{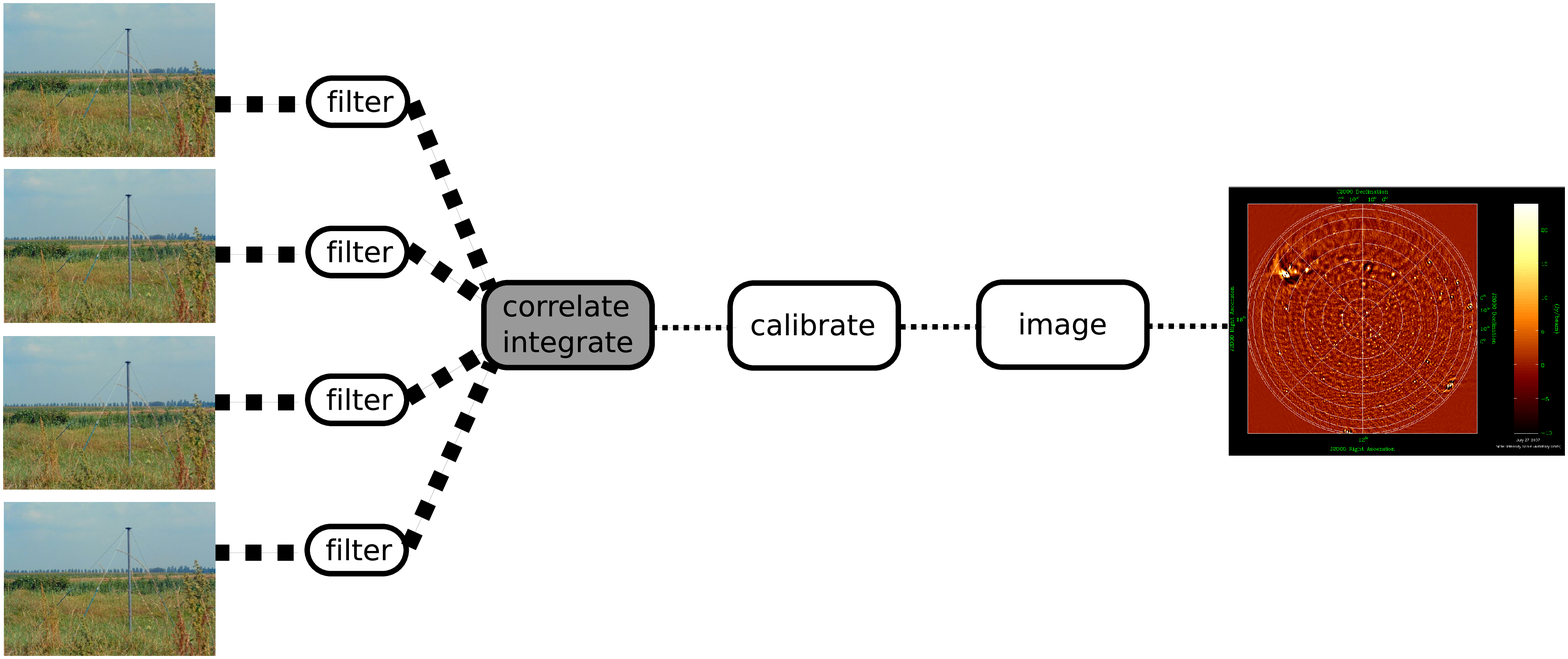}
\end{center}
\caption{An extremely simplified view of LOFAR processing.}
\label{fig-processing-overview}
\end{figure*}

In this paper, we investigate the correlation algorithm on many-core
hardware, such as graphics processors (GPUs)~\cite{gpgpu-hardware} and
the \mbox{Cell/B.E.}~\cite{cell}.  In contrast to many others, we do not
only use NVIDIA GPUs, but also include ATI hardware.  In addition, we
compare with the LOFAR production implementation on a Blue Gene/P
supercomputer~\cite{ppopp2010}.  As a reference, we also include 
multi-core general-purpose processors.  There are
many advantages to the use of many-core systems: it is a flexible
software solution, has lower costs in terms of purchase and maintenance, 
and the power usage is significantly lower than that of a traditional supercomputer.

The correlator differs from applications that were
investigated on many-core hardware in the past, because of the correlator's low flop/byte
ratio.  In addition, it is a streaming real-time application, so host-to-device
data transfers are on the critical path. In many other studies, these
transfers are not considered. 

The production correlator on the Blue Gene/P achieves 96\% of the
theoretical peak performance.  We demonstrate that the processing
power and memory bandwidth of current GPUs are highly imbalanced for
correlation purposes.  This leads to suboptimal performance. Still, GPUs are
considerably more power efficient than the BG/P (4x for NVIDIA, and 2.4x for ATI).
\shortversion{
The \mbox{Cell/B.E.} processor, in contrast, achieves an excellent 92\% efficiency. The \mbox{Cell/B.E.} 
runs the correlator 4-6 times (depending on the manufacturing process of the \mbox{Cell/B.E.}) more energy
efficiently than the Blue Gene/P.
The research presented in this paper is an
important pathfinder for next-generation telescopes.
}
\longversion{
The
correlator achieves only only 8\% of the peak performance on ATI GPUs,
and 18\% on NVIDIA GPUs. 

The \mbox{Cell/B.E.} processor, in contrast,
achieves an excellent 92\% efficiency. The \mbox{Cell/B.E.} is also the most
energy-efficient solution. It runs the correlator 5 times more energy
efficiently than the Blue Gene/P.
The research presented in this paper is an
important pathfinder for next-generation telescopes.
}

The rest of this paper is structured as follows.
Section~\ref{correlation-alg} explains how the correlation algorithm
works, and why it is important. In Section~\ref{hardware}, we describe
the many-core architectures that we evaluate in detail, finishing
with a comparison and discussion.  Next, in Section~\ref{sec:perf}, we
explain how we implemented the correlator algorithm on each of these
architectures, and describe the performance we achieve. In
Section~\ref{sec:perf-compare}, we evaluate, compare, and discuss the
results, while identifying the weak and strong points of the
architectures.  Section~\ref{related} discusses related work.  In
Section~\ref{sec:discussion}, we investigate if our results and
insights can be applied to other applications. Additionally, we discuss scalability
issues. Finally, we conclude in Section~\ref{conclusions}.

\section{Correlating Radio Astronomy Signals}
\label{correlation-alg}


We call a set of receivers that are grouped closely
together a \emph{station}. The data streams from the different
stations must be filtered, delays in the signal path must be
compensated for, and the data streams from different stations must be
cross-correlated. The correlation process performs a data reduction by
integrating samples over time. In this paper, we use the LOFAR
telescope as an example, but the results apply equally well
to other instruments. An overview of the processing needed for the
standard imaging pipeline of LOFAR is shown in
Figure~\ref{fig-processing-overview}. The pipeline runs from left to right.
The thickness of the lines indicates the size of the data streams.
In this paper, we focus on the
correlator step (the gray box in
Figure~\ref{fig-processing-overview}), because its costs grow
quadratically with the number of stations. All other steps have a
lower time complexity. We choose 64 as the number of stations, since
that is a realistic number for LOFAR.  Future instruments will likely
have even more stations. We call the combination of two stations a
\emph{baseline}.  The total number of baselines is $(nrStations \times
(nrStations + 1)) / 2$, since we need each pair of correlations only
once. This includes the autocorrelations (the correlation of a station with itself),
since we need this later in the pipeline for calibration purposes.
Although the autocorrelations can be computed with fewer instructions, we 
ignore this here, since the number of autocorrelations is small, and grows
linearly with the number of stations, while the number of normal correlations
grows quadratically.

The correlator algorithm itself is straightforward, and can be
written in a single formula:
$C_{s_1,s_2\geq s_1,p_1\in\{X,Y\},p_2\in\{X,Y\}} = \displaystyle\sum_{t} Z_{s_1,t,p_1} * Z_{s_2,t,p_2}^\ast$ 

\begin{figure*}[t]
\lstset{language=C,basicstyle={\fontencoding{T1}\fontfamily{pcr}\fontseries{m}\fontshape{n}\fontsize{7}{10pt}\selectfont}} 
\begin{lstlisting}{}
for (ch=0; ch<nrChannels; ch++)
  for (station2=0; station2<nrStations; station2++)
    for (station1=0; station1<=station2; station1++)
      for (pol1 = 0; pol1 < nrPolarizations; pol1++)
        for (pol2 = 0; pol2 < nrPolarizations; pol2++) {
          complex float sum = 0 + i*0;
          for (time=0; time < integrationTime; time++) {
            sum +=  samples[ch][station1][time][pol1]
                 * ~samples[ch][station2][time][pol2];
          }
          baseline = computeBaseline(station1, station2);
          correlation[baseline][ch][pol1][pol2] = sum;
        }
\end{lstlisting}
\caption{Pseudo code for the correlation algorithm.}
\label{correlator-code}
\end{figure*}

Pseudo code for the algorithm is shown in Figure~\ref{correlator-code}.
A sample is a ($2 \times 32-bit$) complex number that represents the
amplitude and phase of a signal at a particular time. The receivers
are polarized; they take separate samples from orthogonal (X and Y)
directions. The received signals from sky sources are so weak, that the antennas 
mainly receive noise. To see if there is statistical coherence
in the noise, simultaneous samples of each pair of stations are correlated, 
by multiplying the sample of one station with the complex
conjugate (i.e., the imaginary part is negated) of the sample of the other station.
To reduce the output size, the products are integrated, by accumulating all products. 
For the LOFAR telescope, we accumulate 768 correlations at 763 Hz, 
so that the integration time is approximately one second. 
This is much shorter than for current telescopes. The short integration time
leads to more output data. 
Since the correlation of station A and B is
the complex conjugate of the correlation of station B and A, only
one pair is computed. 
Stations are also autocorrelated, i.e., with
themselves. Both polarizations of a station A are correlated with both polarizations 
of a station B, yielding correlations in XX, XY, YX, and YY
directions.
The correlator is mostly multiplying and adding complex numbers.
\longversion{
A complex multiplication of two complex numbers $a$ and $b$ can be written as follows, where $a_r$ is the real part of
$a$, while $a_i$ is the imaginary part: \\

\noindent $(a_r + a_ii)(b_r + b_ii)$ \\

\noindent For a correlation, we need to multiply a sample from station A with
the complex conjugate (i.e., the imaginary part is negated) from
station B, and add it to an accumulator to integrate over time. 
}
\longversion{.

\noindent $acc = acc + (a_r + a_ii)(b_r - b_ii)$ \\

\noindent This the same as: \\

\noindent $acc = acc + a_r b_r - a_r b_ii + a_ii b_r - a_ii b_ii$ \\

\noindent The last term ($- a_ii b_ii$) is the same as $+ a_i b_i$, since $i^2 = -1$.
Thus, the total term becomes: \\

\noindent $acc = acc + (a_r b_r + a_i b_i) + (-a_r b_i + a_i b_r )i$ \\

\noindent If we split the real and imaginary parts, and add them to the corresponding accumulators, we get: \\

\noindent $acc_r = acc_r + a_rb_r + a_ib_i$ \\
\noindent $acc_i = acc_i - a_rb_i + a_ib_r$ \\

\noindent Thus, we use both the real and imaginary parts of each sample \emph{two} times.
With the use of fused-multiply add instructions (fma), this becomes: \\
}



We can implement the correlation operation very efficiently, with only
four fma instructions, doing eight floating-point operations in
total. For each pair of stations, we have to do this four times, once
for each combination of polarizations. Thus, in total we need 32
operations and load 8 floats (32 bytes) from memory, resulting in \emph{exactly
  one FLOP/byte}.  The number of operations that is performed per byte
that has to be loaded from main memory is called the \emph{arithmetic
  intensity}~\cite{system-performance}. For the correlation algorithm,
the arithmetic intensity is extremely low.

\begin{figure}[t]
\begin{center}
\includegraphics[width=4.2cm]{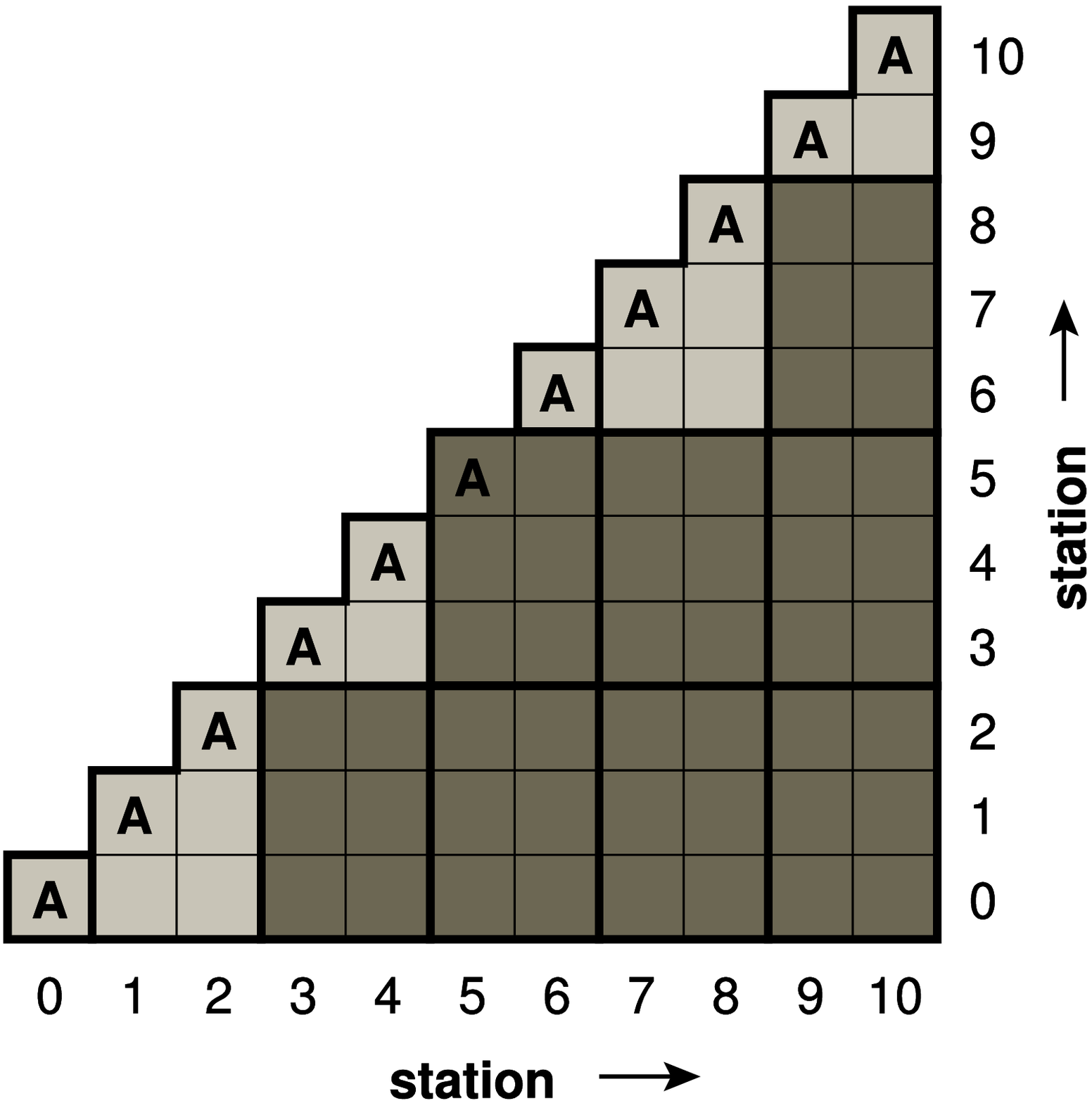}
\end{center}
\caption{An example correlation triangle.}
\label{fig-correlation}
\end{figure}

An important optimization that we implemented is the reduction
of memory loads by the correlator. 
This is achieved
by keeping correlations that are being accumulated in registers, and
by reusing samples that are loaded from memory as many times as
possible. 
A sample can be used multiple times by correlating it
with the samples from multiple other stations in the same loop iteration.
For example, a sample from station A in the X polarization
that is loaded into a register pair can be correlated with the X and
Y polarizations of stations B, C and D, using it 6 times. 
\longversion{
In fact,
it is used 12 times, since a correlation requires 2 complex, fused
multiply-add instructions. 
}
Figure~\ref{fig-correlation} shows how we correlate multiple
stations at the same time. Each square represents the XX, XY,
YX, and YY correlations of the stations as indicated by row and
column number. The figure is triangular, because we compute
the correlation of each pair of stations only once. The squares labeled \emph{A} are
autocorrelations, which could be treated specially since they require less
computations. The triangle is divided into larger tiles, in this case 
2x3 tiles (the dark gray boxes), but arbitrary sizes are possible.
A tile is correlated as a unit. For example, the lower
right-hand-side rectangle correlates stations 9 and 10 with stations
0, 1, and 2.

It is important to tune the tile size to the architecture. We want to
make the tile size as large as possible, while still fitting in the
register file. This offers the highest level of data reuse.  
If we have a $w \times h$ tile size, the number of operations is given by $flops = 32wh$.
The number of bytes that has to loaded from memory is $16(w+h)$.
The minimum number of registers that is required is $4 (1 + min(w,h)) + 8 w h$.
This is the total number of registers, including accumulators, while reusing
registers if a value is no longer needed (hence the $min$ operation). However,
this formula does not count additional registers that could be needed for data prefetching,
address calculations and loop counters.
The number of registers is expressed in single-precision float registers. If an architecture has vector
registers, the result can be divided by the vector length.
Table~\ref{tile-size-table} shows the properties of different tile sizes. 

Despite the division of the correlation triangle in tiles, there
still is opportunity for additional data reuse \emph{between} tiles. 
The tiles
within a row or column in the triangle still need the same samples.
In
addition to registers, caches can thus also be used to increase data
reuse.  Since we know exactly what data can be reused at what moment, we
found it is important to have direct influence on the caches and the thread scheduler.  This
way, we can make sure that tiles in same row or column are calculated
at the same time by different threads. 
Because the algorithm is
extremely data intensive, the resulting optimized implementation on
many-cores is typically limited by the architecture's memory
bandwidth. The memory aspects of the algorithm are twofold.
There is an algorithmic part, the tile size, which is limited
by the number of registers. The second aspect is architectural in nature: the cache
sizes, cache hierarchy and hit ratio. Together, these two aspects dictate the
memory bandwidth that is needed to keep the ALUs busy.

\begin{table}
\caption{Properties of different tile sizes.}
\label{tile-size-table}
{\small
\begin{tabular}{l|r|r|r|r}
tile & floating point & memory loads & arithmetic     &  minimum nr.           \\
size & operations     & (bytes)      & intensity      &  registers (floats)    \\
\hline
1x1  &  32            &   32         &   1.00         &  16                    \\
1x2  &  64            &   48         &   1.33         &  24                    \\
1x4  & 128            &   80         &   1.60         &  40                    \\
2x2  & 128            &   64         &   2.00         &  44                    \\
3x2  & 192            &   80         &   2.40         &  60                    \\
3x3  & 288            &   96         &   3.00         &  88                    \\
4x3  & 384            &  112         &   3.43         & 112                    \\
4x4  & 512            &  128         &   4.00         & 148                    \\
\end{tabular}
} 
\end{table}

In this paper, we  focus on the maximal performance that can be achieved
with a \emph{single many-core chip}.  It is important to realize that the
correlator itself is \emph{trivially parallel}, since tens of thousands of
frequency channels can be processed independently.  This allows us to
efficiently exploit many-core hardware. We use floating point instead
of integer operations, since all architectures support this
well. Single precision floating point is accurate enough for our
purposes.

Since the code is relatively straightforward, we implemented the
performance-critical kernel in assembly on all
architectures. Therefore, this paper really compares the hardware
architectures; \emph{compilers do not influence the
  performance}. Although we wrote the critical parts in assembly, the
additional code was written in the natural programming model for each
architecture. Therefore, we also take programmability into
account.
We do this both for the assembly parts, and the high-level code.
The first is a measure of the programmability of the hardware architecture itself.
The second gives an indication of the quality of the software stack.

\section{Many-Core Hardware}
\label{hardware}


Recent many-core architectures present enormous computational
performance at very low costs. For example, the recently introduced
ATI 4870x2 GPU has 1600 compute cores, and achieves a theoretical
performance of 2.4 teraflops, all on a single PCI card. This GPU costs
less than 500 dollars, resulting in an unprecedented price/performance
ratio. The number of cores in the accelerators also increases rapidly. In only a
few years, the number of cores has increased from about 16, e.g., in
the NVIDIA GeForce~6 series in 2004, to the 1600 cores in the ATI
4870x2, that became available in 2008. In only 4 years, a 100-fold
increase in the number of cores has been realized. During this time,
the theoretical peak performance increased from 12~gigaflops to 
2.4~teraflops, a growth of a factor 200. 

Reductions in power costs become increasingly
important. For LOFAR, for example, a large part of the operational
costs is the electrical power consumption. Current supercomputers
already focus on power efficiency. The IBM Blue Gene/P, for
instance, can perform 384 gflops/kW~\cite{bgp}, which is significantly better
than current general-purpose systems. Nevertheless, modern many-core
accelerators are, in theory, even more efficient. For example, an NVIDIA GTX280
GPU has a theoretical peak performance of 3952 gflops/kW, more than ten times
better than a Blue Gene/P. The IBM Roadrunner system~\cite{roadrunner}, which is based
on the \mbox{Cell/B.E.} processor, became the first super computer to achieve one
petaflops. The three most energy efficient supercomputers on the
Green500 list\footnote{See http://www.green500.org.}, are similarly based on the \mbox{Cell/B.E.} 

Although the number of cores and the theoretical peak performance have
increased dramatically, the memory bandwidth did not increase equally
fast.  In fact, the available memory bandwidth \emph{per core} is
decreasing for many-core architectures at the moment.  This introduces
significant performance bottlenecks for data intensive
applications~\cite{gpgpu-hardware,larrabee}.  The important question
is: How much of the theoretical performance can be reached in
practice? In this paper we answer this question for correlation
process, we extrapolate our results to data-intensive applications in
general.  The increase of both the number of cores and the peak
performance is much larger for the accelerators than for normal
CPUs. This makes many-core accelerators interesting for
high-performance computing.  In the remainder of this section, we
discuss several many-core architectures in detail, and conclude with a
summary and discussion of the differences that are essential for the
correlator, and for data-intensive applications in general.

\subsection{General Purpose multi-core CPU (Intel Core i7 920)}

As a reference, we implemented the correlator on a multi-core general
purpose architecture. We use a quad core Intel Core~i7 920 CPU
(code name Nehalem) at 2.67~GHz. 
\longversion{
The chip is manufactured with a 45~nm 
process.  The on-chip L1 cache has 32 KB for
instructions, and 32~KB for data per core. 

There is 256 KB L2 cache (combined instruction and data) per core, and
8 MB L3, shared by all cores.
The system uses a 4.8 GT/s QuickPath interface, and has a thermal design power of 80~Watts. 
}
\shortversion{
There is 32~KB of on-chip L1 data cache per core, 256~KB L2 cache per core, and 8~MB
of shared L3 cache.  
The thermal design power (TDP) is 130~Watts.
}
The theoretical
peak performance of the system is 85~gflops, in single precision.
The parallelism comes from four cores with two-way hyperthreading, and a vector length of four floats,
provided by the SSE4 instruction set.  

The architecture has several
important drawbacks for our application.  First, there is no fused
multiply-add instruction.  Since the correlator performs mostly
multiplies and adds, this can cause a performance penalty. The
processor does have multiple pipelines, and the multiply and add
instructions are executed in different pipelines, allowing eight
flops per cycle per core.

Another problem is that SSE's shuffle instructions
to move data around in vector registers are more limited
than for instance on the \mbox{Cell/B.E.} processor. This complicates an
efficient implementation.
For the future Intel Larrabee GPU, and for the next
generation of Intel processors, both a fused multiply-add instruction
and improved shuffle support has been announced.

The number
of SSE registers is small (sixteen 128-bit registers), allowing only little
data reuse. 
\longversion{At most 64 single-precision floats can be kept in registers.}
This is a problem for the correlator, since
the tile size is limited by the number of registers.  A smaller tile
size means less opportunity for data reuse, increasing the memory
bandwidth that is required.






\subsection{IBM Blue Gene/P}

The IBM Blue Gene/P~(BG/P)~\cite{bgp} is the architecture that is
currently used for the LOFAR correlator~\cite{ppopp2010}.  Four 850~MHz
PowerPC~450 processors are integrated on each Blue Gene/P chip.  We
found that the BG/P is extremely suitable for our application, since
it is highly optimized for processing of complex numbers.  The BG/P
performs \emph{all} floating point operations in double precision, which is
overkill for our application.
The L2 prefetch unit prefetches the sample data efficiently from
memory.  In contrast to all other architectures we evaluate, the
problem is compute bound instead of I/O bound, thanks to the BG/P's high
memory bandwidth per operation. It is 3.5--10 times higher than
for the other architectures.  
The ratio between flops and bytes/sec of
memory bandwidth is exactly 1.0 for the BG/P.

The BG/P has a register file with 32 vector registers of width 2.
Therefore, 64 floating point numbers (with double precision) can be
kept in the register file simultaneously. This is the same amount as
on the general purpose Intel chip, but an important difference is that the
BG/P has 32 registers of width 2, compared to Intel's 16 of width 4.
The smaller vector size reduces the amount of shuffle instructions
needed. 
\longversion{
 Also, the available memory bandwidth per operation is much
higher on the Blue Gene/P, so this is not a problem.
}
The BG/P is an energy efficient supercomputer. This is
accomplished by using many small, low-power chips, at a low clock
frequency.  The supercomputer also has excellent I/O capabilities,
there are five specialized networks for communication.

\subsection{ATI 4870 GPU (RV 770)}

\shortversion{
The most high-end GPU provided by ATI (recently acquired by AMD) is the 4870~\cite{amd-manual}. 
The RV770 processor in the 4870 runs at 750 MHz, and has a thermal design
power of 160 Watts. 
The RV770 chip has ten SIMD cores, each containing 16 
superscalar streaming processors. Each streaming
processor has five independent scalar ALUs. Therefore, the GPU
contains 800 ($10 \times 16 \times 5$) scalar 32-bit streaming processors.  The
Ultra-Threaded Dispatch Processor controls how the execution units
process streams. 
The theoretical peak performance is 1.2~teraflops.
The 4870 has 1~GB of GDDR5 memory with a theoretical bandwidth of 115.2~GB/s.
The board uses a PCI-express~2.0 interface for communication with
the host system.
Each of the ten SIMD cores contains 16 KB of local memory and separate L1
texture cache.
The L2 cache is shared. The
maximum L1 bandwidth is 480 GB/sec. The bandwidth between the L1 and
L2 Caches is 384 GB/sec. The application can specify if a read should be cached or not.
The SIMD cores can exchange data using 16 KB of global
memory.
}

The ATI 4870 GPU has the largest number of cores of all architectures
we evaluate (800).  However, the architecture has several important
drawbacks for data-intensive applications.  First, the
host-to-device bandwidth is too low. In practice, the achieved
PCI-express bandwidth is far from the theoretical limit.  We will
explain this in more detail in Section~\ref{perf-ati}.  The achieved
bandwidth is not enough to keep all cores busy.  Second, we found that
overlapping communication with computation by performing asynchronous
data transfers between the host and the device has a large impact on
kernel performance. We observed kernel slowdowns of \emph{a factor of
  three} due to transfers in the background.  Third, the architecture
does not provide random write access to device memory, but only to
\emph{host} memory. However, for our application which is mostly
read-performance bound, this does not have a large impact (see
Section~\ref{perf-ati}).



\subsection{NVIDIA GPU (Tesla C1060)}

\shortversion{
NVIDIA's Tesla C1060 contains a GTX~280 GPU (code-named GT200), is
manufactured using a 65 nm process, and has 1.4 billion
transistors. The device has 30 cores (called multiprocessors) running
at 1296 MHz, with 8 single precision ALUs, and one double precision
ALU per core.  Current NVIDIA GPUs thus have fewer cores than ATI
GPUs, but the individual cores are faster. The memory architecture is
also quite different. NVIDIA GPUs still use GDDR3 memory, while ATI
already uses GDDR5 with the 4870~GPU. The GTX~280 in the Tesla
configuration has 4~GB of device memory, and has a thermal design
power of 236 Watts.  The theoretical peak performance is 933 gflops.

The number of registers is large: there are 16384 32-bit floating
point registers per multiprocessor. There also is 16~KB of shared memory per multiprocessor. 
This memory is shared between all threads on a multiprocessor, but not globally.
There is a total amount of 64 KB of constant memory on the chip.  
Finally, texture caching hardware is available. 
NVIDIA only specifies that ``the cache working set for texture memory
is between 6 and 8 KB per multiprocessor''~\cite{cuda-manual}.
The application has some control over the caching
hardware.  It is possible to specify which area of device
memory must be cached, while the shared memory is completely
managed by the application.  

On GPUs, it is possible to synchronize the threads within a multiprocessor.
With our application, we exploit this to increase the cache hit
ratio. This improves performance considerably on NVIDIA hardware, but not on ATI hardware. 
When accessing device memory, it is important to make sure that simultaneous
memory accesses by different threads are \emph{coalesced} into a
single memory transaction.  
In contrast to ATI hardware, NVIDIA GPUs support random write access
to device memory. This allows a programming model that is much closer
to traditional models, greatly simplifying software development.
The NVIDIA GPUs suffer from a similar 
problem as the ATI GPUs: the host-to-device bandwidth is equally
low.
}

\begin{table}
\caption{Differences between many-core memory architectures.}
\label{memory-properties}
{\small
\begin{tabular}{l|l|l}
feature                   & Cell/B.E.                              & GPUs \\
\hline
\hline
access times              & uniform                                & non-uniform \\
\hline
cache sharing level       & single thread (SPE)                    & all threads in a \\
                          &                                        & multiprocessor \\
\hline
access to off-chip memory & not possible, only through DMA         & supported \\
\hline
memory access             & asynchronous DMA                       & hardware-managed \\
overlapping               &                                        & thread preemption \\
\hline
communication             & communication between                  & independent thread blocks \\
                          & SPEs through EIB                       & + shared memory within a block \\
\end{tabular}
} 
\end{table}

\subsection{The Cell Broadband Engine (QS21 blade server)}

\shortversion{
The Cell Broadband Engine (\mbox{Cell/B.E.})~\cite{cell} is a heterogeneous many-core
processor, designed by Sony, Toshiba and IBM (STI). 
The \mbox{Cell/B.E.} has nine cores: the Power Processing Element
(PPE), acting as a main processor, and eight Synergistic Processing
Elements (SPEs) that provide the real processing power. All cores run at 3.2 GHz.
The cores, the main memory, and the external I/O are connected by a
high-bandwidth (205 GB/s) Element Interconnection Bus (EIB).
The main memory has a high-bandwidth (25 GB/s), and uses XDR (Rambus).
The PPE's main role is
to run the operating system and to coordinate the SPEs.
An SPE contains a RISC-core (the Synergistic Processing Unit (SPU)), 
a 256KB Local Store (LS), and a memory flow controller. 

The LS is an extremely fast local
memory (SRAM) for both code and data and is managed entirely by the
application with explicit DMA transfers.  The LS can be considered 
the SPU's L1 cache.  
The LS bandwidth is 47.7 GB/s per SPU.
The \mbox{Cell/B.E.} has a large number of registers: each SPU has 128, which are
128-bit (4 floats) wide. The theoretical peak performance of one SPU
is 25.6 single-precision gflops.
The SPU can dispatch two instructions in each clock cycle using 
the two pipelines designated \emph{even} and
\emph{odd}. Most of the arithmetic instructions execute on the even
pipe, while most of the memory instructions execute on the odd pipe.
We use a QS21 Cell blade with two \mbox{Cell/B.E.} processors and 2 GB
main memory (XDR). This is divided into 1 GB per processor.  
A single \mbox{Cell/B.E.} in our system has a TDP of 70~W. 
Recently, an equally fast version with a 50~W TDP has been announced.
The 8 SPEs of a single chip in the system have a total theoretical single-precision peak performance of 205 gflops.
}

\subsection{Hardware Comparison and Discussion}
\label{hardware-comparision}

The memory architectures of the many-core systems are of particular
interest, since our application is mostly memory-throughput 
bound (as will be discussed in Section~\ref{sec:perf}).
Table~\ref{memory-properties} shows some key differences of the memory
architectures of the many-core systems.
\longversion{
The caching
architectures of the many-core systems show several remarkable
differences.
}
Both ATI and NVIDIA GPUs have a hardware L1 and L2
cache, where the application can control which memory area is cached,
and which is not.  The GPUs also have shared memory, which is
completely managed by the application.  
\longversion{
This memory is shared between
threads in the same block, but not globally between blocks. 
}
Also,
coalescing and bank conflicts have to be taken into account, at the
cost of significant performance penalties~\cite{cuda-manual}.  Therefore, the
memory access times are \emph{non-uniform}.  The access
times of the local store of the \mbox{Cell/B.E.}, in contrast, are completely
uniform (6 cycles).  Also, each \mbox{Cell/B.E.} SPE has its own \emph{private} local store, there
is no cache that is shared between threads.  While the GPUs can
directly access device memory, the \mbox{Cell/B.E.} does not provide access to
main memory. All data has to be loaded and stored into the local store
first. Also, the way that is used to overlap memory accesses
with computations is different.  The \mbox{Cell/B.E.} uses
asynchronous DMA transfers, while the GPUs use hardware-managed thread
preemption to hide load delays.  Finally, the SPEs of the \mbox{Cell/B.E.} can
communicate using the Element Interconnection Bus, while the multiprocessors
of a GPU execute completely independently.

\begin{table}[t]
\caption{Properties of the different many-core hardware platforms. For the Cell/B.E., we consider the local store to be L1 cache.}
\label{architecture-properties}
{\small
\begin{tabular}{l|l|l|l|l|l}                                                   
                                             & Intel         & IBM            & ATI           &  NVIDIA       & STI          \\
Architecture                                 &       Core i7 & BG/P           &     4870      &   Tesla C1060 &     Cell/B.E.\\
\hline
cores x FPUs per core                        & 4x4           & 4x2            & 160x5         & 30x8          & 8x4          \\
operations per cycle per FPU                 & 2             &   2            & 2             & 2             & 2            \\
Clock frequency (GHz)                        & 2.67          & 0.850          & 0.75          & 1.296         & 3.2          \\
\textbf{gflops per chip}                     & \textbf{85}   & \textbf{13.6}  & \textbf{1200} & \textbf{936}  & \textbf{204.8}\\
\hline
registers per core x register width          & 16x4          & 64x2           & 1024x4        & 2048x1         & 128x4        \\
\hline
total L1 data cache size per chip (KB)       & 32            & 128            & ???           & ???            & 2048         \\
total L1 cache bandwidth (GB/s)              & ???           & 54.4           & 480           & ???            & 409.6        \\
total device RAM bandwidth (GB/s)            & n.a.          & n.a.           & 115.2         & 102            & n.a.         \\
\textbf{total host RAM bandwidth (GB/s)}     & \textbf{25.6} & \textbf{13.6}  & \textbf{8.0}  & \textbf{8.0}   & \textbf{25.8}\\
\hline
Process Technology (nm)                      & 45            & 90             & 55            & 65             & 65           \\
TDP (W)                                      & 130           & 24             & 160           & 236            & 70           \\
\textbf{gflops / Watt (based on TDP)}        & \textbf{0.65} & \textbf{0.57}  & \textbf{7.50} & \textbf{3.97}  & \textbf{2.93}\\
\hline
\textbf{gflops/device bandwidth (gflops / GB/s)}& n.a.       &  n.a.          & \textbf{10.4} & \textbf{9.2}   & n.a.         \\
\textbf{gflops/host bandwidth (gflops / GB/s)} & \textbf{3.3}& \textbf{1.0}   & \textbf{150}  & \textbf{117}   & \textbf{7.9} \\
\end{tabular}
} 
\end{table}


Table~\ref{architecture-properties} shows the key properties of the
different architectures we discuss here. Note that the performance
numbers indicate the \emph{theoretical} peak.  The memory bandwidths of the
different architectures show large differences.  Due to the PCI-e bus,
the host-to-device bandwidth of the GPUs is low. 
The number of gflops per byte of memory bandwidth gives an indication of the performance of the 
memory system. A lower number means a better balance between memory
and compute performance.  For the GPUs, we can split this number into
a device-to-host component and an internal component. It is clear that
the \emph{relative} performance of the memory system in the Blue
Gene/P system is significant higher than that of all the other
architectures. 
The number of gflops that can be achieved per Watt is
an indication of the theoretical power efficiency of the hardware.  
In theory, the many-core architectures are more power efficient than
general-purpose systems and the BG/P.

The Bound and Bottleneck analysis~\cite{system-performance,roofline} 
is a method to gain insight into
the performance that can be achieved in practice on a particular platform. 
Performance is bound \emph{both}
by theoretical peak performance in flops, and the product of the
memory bandwidth and the arithmetic intensity $AI$ (the flop/byte
ratio): \\
$\mathit{perf_{max} = min(perf_{peak}, AI \times memoryBandwidth)}$.
Several important assumptions are made with this method. First, it
assumes that the memory bandwidth is independent of the access
pattern.  Second, it assumes a complete overlap of communication and
communication, i.e., all memory latencies are completely
hidden.  Finally, the method does not take caches into
account. Therefore, if the correlator can make effective use of the
caching mechanisms, performance can actually be better than
$\mathit{perf_{max}}$. Nevertheless, the $\mathit{perf_{max}}$ gives a rough idea of the
performance than can be achieved.  

With the GPUs, there are several communication steps that influence
the performance. First, the data has to be transferred from the host to
the device memory.  Next, the data is read from the device memory into
registers. 
\shortversion{
Although the GPUs offer high internal memory bandwidths,
the host-to-device bandwidth is limited by the low
PCI-express throughput (8~GB/s for PCI-e~2.0~16X). 
In practice, we measured even lower throughputs.
With the NVIDIA GPU, we achieved 5.58~GB/s, and with the ATI GPU 4.62 GB/s.
}
\longversion{
Although the GPUs offer high internal memory bandwidths,
(115.2~GB/s for the ATI~4870, and 141.7~GB/s for the NVIDIA~GTX280)
the bandwidth from the host to the device is limited by the low
PCI-express bandwidth (8~GB/s for PCI-e~2.0~16X). 
Therefore, we
investigate the performance that can be achieved both with and without
communication from the host to the device.
In practice, we measured even lower throughputs from the host to the GPUs.
With the NVIDIA GPU, we achieved 5.58~GB/s, and with the ATI GPU 4.62 GB/s.
The throughputs of the different GPUs thus is almost identical, and much lower than
the theoretical limit.
}
For both communication steps, we can compute the arithmetic intensity and the
$\mathit{perf_{max}}$. The sample data must be loaded into the device memory, but
is then reused several times, by the different tiles. We call the 
arithmetic intensity from the point of view of the entire computation $AI_{global}$.
The number of flops in the computation is the number of baselines times 32 operations,
while the number of bytes that have to be loaded in total is 16 bytes times the number of stations.
As explained in Section~\ref{correlation-alg}, the number of baselines is $(nrStations \times (nrStations + 1)) / 2$.
If we substitute this, we find that $AI_{global} = nrStations + 1$.
Since we use 64 stations, the $AI_{global}$ is 65 in our case.
The $AI_{local}$ is the arithmetic intensity on the device itself. The value depends on the tile size,
and was described in Section~\ref{correlation-alg}.

For both ATI and NVIDIA hardware, the $\mathit{perf_{max,global}}$ is $65 \times 8.0 =
520$ gflops, if we use the theoretical PCI-e~2.0~16X bandwidth of 8~GB/s.
If
we look at the PCI-e bandwidth that is achieved in practice (4.62 and 5.58 GB/s respectively), 
the GPUs have a $\mathit{perf_{max,global}}$ of \emph{only 300 gflops for ATI, and 363 gflops for NVIDIA}.
Since there is no data reuse
between the computations of different frequency channels, this is a
realistic upper bound for the performance that can be achieved, assuming there is no performance penalty for
overlapping device computation with the host-to-device transfers.  We
conclude that due to the low PCI-e bandwidth, only a small fraction of
the theoretical peak performance can be reached, even if the kernel
itself has ideal performance.
In the following sections we will
evaluate the performance we achieve with the correlator in detail,
while comparing to $\mathit{perf_{max}}$.



\section{Correlator Implementation and Performance}
\label{sec:perf}

This section describes the implementation of the correlator on
the different architectures. 
In many cases, we experimented with
different versions, because it is often unclear beforehand what the
best implementation strategy is.
We evaluate the performance in detail. For comparison reasons, we use the performance
\emph{per chip} for each architecture.
We also calculate the achieved memory bandwidths for all
architectures in the same way.  We know the number of bytes that has
to be loaded by the kernel, depending on the tile size that is used.
We divide this by the
execution time of the kernel to calculate the bandwidth. 
Thanks to data reuse with caches and local stores, the achieved bandwidth can be \emph{higher}
than the memory bandwidth.

\subsection{General Purpose multi-core CPU (Intel Core i7 920)}

\longversion{
\begin{table}[t]
\begin{center}
\begin{tabular}{l||r|r||r|r}
& \multicolumn{2}{c||}{1 core        } &\multicolumn{2}{c}{4 cores} \\
tile &                & throughput     &               & throughput     \\
size & gflops         & (GB/s)         & gflops        & (GB/s)         \\
\hline
1x1  &  9.2 (51.2\%)  & 8.6  (33.6\%) & 35.7 (49.6\%) & 33.3 (130.1\%) \\
2x2  & 11.1 (61.5\%)  & 5.2  (20.3\%) & 44.3 (61.5\%) & 20.6 ( 80.5\%) \\
3x2  & 10.7 (59.7\%)  & 4.1  (16.0\%) & 43.0 (59.7\%) & 16.7 ( 65.2\%) \\
\end{tabular}
\end{center}
\caption{Performance of the correlator on a core i7 920.}
\label{results-cell}
\end{table}
}

We use the SSE3 instruction set to exploit vector parallelism. 
Due to the limited shuffle instructions, computing the correlations of the four
polarizations within a vector is inefficient. We achieve only a speedup of a factor
of 2.8 compared to a version without SSE3. 
We found that, unlike on all
other platforms, vectorizing the integration time loop works significantly better.
This way, we compute four samples with subsequent time stamps in a vector.
The use of SSE3 improves the performance by a factor of 3.6 in this case.
In addition, we use multiple threads to utilize all four cores. 
To benefit from hyperthreading, we need twice as many threads as cores (i.e., 8 in our case).
Using more threads does not help. Hyperthreading increases performance by 6\%.
The most efficient version uses a tile size of 
$2 \times 2$\longversion{and explicitly spills registers to the stack}.  Larger tile sizes are
inefficient due to the small SSE3 register file. We achieve a
performance of 48.0 gflops, 67\% of the peak, while using 73\% of the peak bandwidth.


\subsection{IBM Blue Gene/P}
The LOFAR production correlator is implemented on the Blue Gene/P platform.
We use it as the reference for performance comparisons.
\longversion{
For optimal performance, most time-intensive code is written in
assembly, since we could not get satisfactory performance from
compiled C++ code. 
}
The (assembly) code hides load and
instruction latencies, issues concurrent floating point, integer, and
load/store instructions, and uses the L2 prefetch buffers in the most
optimal way. 
\longversion{
Most instructions are parallel fused multiply-adds,
that sustain four operations per cycle.
We optimally exploited the
large, $32 \times 2$ FPU register file. 
}
We use a cell size of $2 \times 2$, since this offers the highest
level of reuse, while still fitting in the register file. 
\longversion{
Each tile
requires 16 complex registers to accumulate the correlations. With 32
complex register available, there are 16 left to load the X and Y
samples from the stations. The correlation of multiple stations in the
same loop iteration also helps to hide the 5-cycle instruction latencies of
the fused multiply-add instructions, since the correlations are
independently computed.
}
The performance we achieve with this version is 13.1 gflops per chip,
96\% of the theoretical peak performance. 
\longversion{
The Blue Gene/P has enough
memory bandwidth to keep all cores busy.
}
The problem is compute bound, and not I/O bound, thanks to the
high memory bandwidth per flop, as is shown in Table~\ref{architecture-properties}.
For more information, we refer to~\cite{ppopp2010}.

\subsection{ATI 4870 GPU (RV 770)}
\label{perf-ati}

ATI offers two separate programming models, at
different abstraction levels.  The low-level programming model is
called the ``Compute Abstraction Layer'' (CAL).  
\longversion{
CAL provides
direct access to GPU without needing to learn graphics-specific
programming languages. 
}
CAL provides communication primitives and an
intermediate assembly language, allowing fine-tuning of
device performance. For
high-level programming, ATI adopted \emph{Brook}, which was originally developed at
Stanford~\cite{brook}. ATI's extended version is called
\emph{Brook+}~\cite{amd-manual}.
We implemented the correlator both with Brook+ and with CAL.

\longversion{
The maximal resolution of textures is $8192\times8192$.  This
hardware limitation is reflected in the programming models that AMD provides:
arrays are also maximally $8192\times8192$ large.  
}

With both Brook+ and CAL, the programmer has to do the vectorization, unlike with NVIDIA GPUs.
CAL provides a feature called \emph{swizzling}, which is used to
select parts of vector registers in arithmetic operations.
 For
example, it is possible to perform operations like \texttt{mul r0,
  r1.xxyy, r2.zwzw}, which selects the $x$ and $y$ components of $r1$
twice, and performs a vector multiplication with the $z$ and $w$
components of $r2$, resulting in the four combinations.
We extensively exploit this feature in the correlator, and 
found that this
improves readability of the code significantly.
Unlike the other architectures, the ATI GPUs are not well documented.
Essential information, such as the number of registers, cache sizes,
and memory architecture is missing, making it hard to write
optimized code. Although the situation improved recently, the documentation is still inadequate.
Moreover, the programming tools are insufficient. The
high-level Brook+ model does not achieve acceptable performance for our application. The
low-level CAL model does, but it is difficult to use. 

\longversion{
\begin{table}[t]
\begin{center}
{\small
\begin{tabular}{l|r||r|r||r|r||r}
& & \multicolumn{2}{c||}{no communication} &\multicolumn{2}{c||}{with communication}   & cache   \\
               & tile &                & throughput    &               & throughput     & hit     \\
implementation & size & gflops         & (GB/s)        & gflops        & (GB/s)         & ratio   \\
\hline
Brook+ regs    & 1x1  &       (  \%) &      (  \%)     &      (   \%) &      (  \%)  & unknown \\ 
Brook+ regs    & 2x2  &       (  \%) &      (  \%)     &      (   \%) &      (  \%)  & unknown \\ 
\hline
CAL regs       & 1x1  &   97  ( 8\%) & 91   (79\%)     &  75  ( 6\%)  & 70   (61\%)  & 23\% \\
CAL regs       & 2x2  &  163  (14\%) & 76   (58\%)     &  88  ( 7\%)  & 50   (33\%)  & 13\% \\
\hline
CAL mem        & 1x1  &   73  ( 6\%) & 68   (59\%)     &  59  ( 5\%)  & 55   (48\%)  & 16\% \\
CAL mem        & 2x2  &  171  (14\%) & 80   (69\%)     & 110  ( 9\%)  & 52   (45\%)  & 15\% \\
CAL mem        & 3x2  &  182  (15\%) & 71   (61\%)     & 113  ( 9\%)  & 44   (38\%)  & 12\% \\
CAL mem        & 3x3  &  296  (25\%) & 92   (80\%)     & 147  (12\%)  & 46   (40\%)  & 27\% \\
CAL mem        & 4x3  &  404  (34\%) &110   (95\%)     & 171  (14\%)  & 47   (41\%)  & 42\% \\
\end{tabular}
} 
\end{center}
\caption{Performance of the correlator on an ATI GPU.}
\label{results-ati}
\end{table}
}

Synchronizning the threads within a multiprocessor can increase the cache hit ratio, by
ensuring that threads that access the same samples are scheduled at roughly
the same time. With NVIDIA hardware, this leads to a considerable
performance improvement (see Section~\ref{nvidia-perf}). 
However, although the ATI hardware can synchronize the threads within a multiprocessor,
we could not achieve performance increases this way.
\longversion{
With ATI
hardware, there is no way to influence the thread scheduler. Moreover,
the way the scheduler works is not documented.
}

\shortversion{
The architecture also does not provide random write access to device
memory. The kernel output can be written to at most 8 output
registers (each 4 floats wide). The hardware stores these to
predetermined locations in device memory.  When using the output
registers, at most 32 floating point values can be stored.  This
effectively limits the tile size to $2\times2$.
Random write access to \emph{host} memory is provided. 
The correlator reduces the data by a large amount, and the results are
never reused by the kernel. Therefore, they can be directly streamed
to host memory.  

The theoretical operations/byte ratio of the ATI 4870 architecture is
10.4 for device memory (see Table~\ref{architecture-properties}).  In
order to achieve this ratio with our application, a minimal tile size
of $10 \times 10$ would be needed.  This would require at least 822
registers per thread. This is unfeasible, so we cannot achieve the
peak performance.  Data sharing between tiles using the hardware
caches could improve this situation.  

The best performing implementation streams the result data directly to
host memory, and uses a tile size of 4x3, thanks to the large number
of registers.  The kernel itself achieves 420 gflops, which is 35\% of
the theoretical peak performance. The achieved device memory bandwidth
is  114~GB/s, which is 99\% of the theoretical maximum. Thanks to the
large tile size, the cache hit ratio is 65\%.  As is shown in
Table~\ref{tile-size-table}, the arithmetic intensity with this tile
size is 3.43.  Therefore, \\
$\mathit{perf_{max} = min(1200, 3.43 \times 115.2) = 395}$. 
We achieve significantly more than this, thanks to the texture cache.

If we also take the host-to-device transfers into account, performance
becomes much worse. 
We found that the host-to-device throughput is only 4.62 GB/s in practice,
although the theoretical PCI-e bus bandwidth is 8 GB/s. 
The transfer can be done asynchronously, overlapping the
computation with host-to-device communication.  However, we discovered that the
performance of the compute kernel decreases significantly
if transfers are performed concurrently. For the $4\times3$
case, the compute kernel becomes 2.2 times slower, which can be fully
attributed to the decrease of device memory throughput.  Due to the
low I/O performance, we achieve only 190 gflops, 16\% of the theoretical
peak. This is 63\% of the $\mathit{perf_{max,global}}$ of 300 gflops
that we calculated in Section~\ref{hardware-comparision}.
}

\subsection{NVIDIA GPU (Tesla C1060)}
\label{nvidia-perf}

NVIDIA's programming model is called Cuda~\cite{cuda-manual}.
Cuda is relatively high-level, and achieves good performance.
However, the programmer still has to think about many details such as
memory coalescing, the texture cache, etc.
An advantage of NVIDIA hardware and Cuda is that the application does not have to do 
vectorization. This is thanks to the fact that all cores have their own address generation units. 
All data parallelism is expressed by using threads.

The correlator uses 128-bit reads to load a complex sample with two
polarizations with one instruction.  Since random write access to
device memory is supported (unlike with the ATI hardware), we can
simply store the output correlations to device memory.  We use the
texture cache to speed-up access to the sample data. We do not use it for the
output data, since that is written only once, and never read back by the kernel. 
With Cuda, threads
within a thread block can be synchronized.  We exploit this feature to let
the threads that access the same samples run in lock step.  This way,
we pay a small synchronization overhead, but we can increase the cache hit
ratio significantly.  We found that this optimization improved performance by a factor of 2.0.

We also investigated the use of the per-multiprocessor shared memory as an
application-managed cache.  Others report good results with this
approach~\cite{gpu-cache}.  However, we found that, for our
application, the use of shared memory only led to performance
degradation.

\longversion{
\begin{table}[t]
\caption{Performance of the correlator on an NVIDIA Tesla GPU.}
{\small
\begin{tabular}{r||r|r||r|r}
     & \multicolumn{2}{c||}{no communication} &\multicolumn{2}{c}{with communication}   \\
tile &                & throughput    &              & throughput     \\
size & gflops         & (GB/s)        & gflops       & (GB/s)         \\
\hline
1x1  & 158 (17\%)     & 147 (144\%)     & 145 (16\%)    & 135 (132\%)  \\
2x2  & 228 (24\%)     & 106 (104\%)     & 202 (22\%)    &  94 ( 92\%)  \\
3x2  & 285 (31\%)     & 110 (108\%)     & 243 (26\%)    &  94 ( 93\%)  \\
3x3  & 280 (30\%)     &  87 ( 85\%)     & 239 (26\%)    &  74 ( 73\%)  \\
4x3  & 101 (11\%)     &  28 ( 27\%)     &  95 (10\%)    &  26 ( 25\%)  \\
\end{tabular}
} 
\label{results-nvidia}
\end{table}
}

\longversion{
As is shown in Table~\ref{results-nvidia}, the best performing
implementation uses a tile size of 3x2.
}
\shortversion{
The best performing
implementation uses a tile size of 3x2.
}
The optimal tile size is influenced by the way the available registers are used.
The register file is a shared resource. A smaller tile size means less register usage, 
which allows the use of more concurrent threads, hiding load delays.
On NVIDIA hardware, we found that the using a relatively small tile size and many threads increases performance.

The kernel itself, without host-to-device communication achieves 314
gflops, which is 34\% of the theoretical peak performance. The
achieved device memory bandwidth is 122~GB/s, which is 120\% of the
theoretical maximum. We can reach more than 100\% because we include
data reuse.  The performance we get with the correlator is
significantly improved thanks to this data reuse, which we achieve by
exploiting the texture cache.  The advantage is large, because
separate bandwidth tests show that the theoretical bandwidth cannot be
reached in practice. Even in the most optimal case, only 71\% (72
GB/s) of the theoretical maximum can be obtained.  The arithmetic
intensity with this tile size is 2.4. We can use this to calculate the
maximal performance without communication.  
$\mathit{perf_{max} = min(966, 2.4 \times 102) = 245}$ gflops.  In practice, the
performance is better than that: we achieve 128\% of this, thanks to
the efficient texture cache.

If we include communication, the performance drops by 15\%, and we
only get 274 gflops. Just like with the ATI hardware, this is caused
by the low PCI-e bandwidth.  With NVIDIA hardware and our
data-intensive kernel, we do see significant performance gains by
using asynchronous I/O. With synchronous I/O, we achieve only 162
gflops (compared to the 274 we get with asynchronous I/O). Therefore,
the use of asynchronous I/O is essential. 

\begin{figure*}[t]
\begin{center}
\includegraphics[width=0.4\columnwidth]{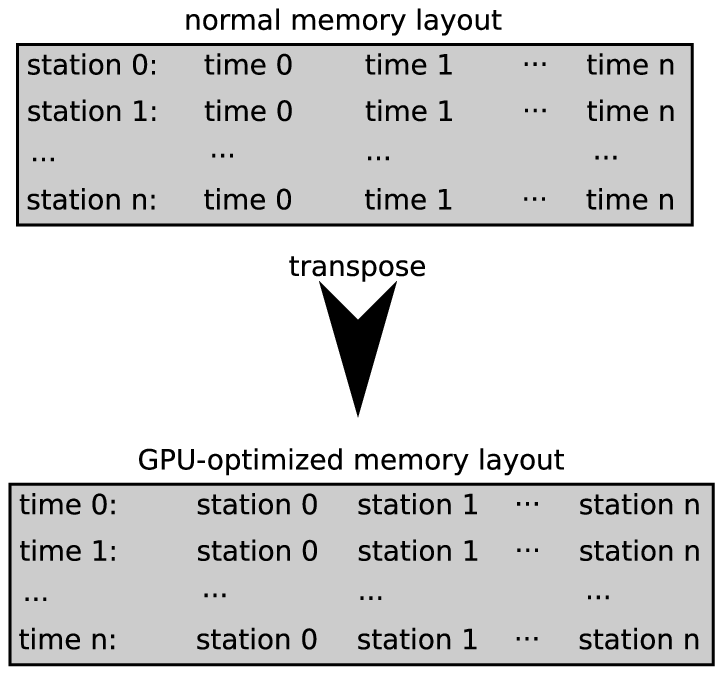}
\end{center}
\caption{Transposing the memory layout for better memory coalescing on the GPU.}
\label{memory-layout}
\end{figure*}

Since memory coalescing is so important on GPUs, we also investigated
an alternative implementation, that first performs a memory transpose.
This way, we can make sure that the GPU threads always read the
samples in a coalesced way as much as possible. Normally, the memory
layout is such that a sequence of samples over time are subsequently
stored in memory per station. For the earlier steps in the correlator
pipeline, this is the natural order. However, if we change this
around, and store the data from all stations for each time step
together, we can achieve better performance on the GPU. The two different
memory layouts are shown in Figure~\ref{memory-layout}.
Of course,
this data transpose is costly.  Nevertheless, we found that the total
performance in the system can be increased, if we perform the
transpose on the host, while the GPU is correlating. Thus, we overlap
the computation on the GPU with both the transpose on the host and
the data transfer from the host to the device. Performing the
transpose on the GPU puts it on the critical path, and decreases
performance considerably.  Our transpose on the host is implemented
using multiple threads to exploit the multi-core architecture, and
using SSE instructions for better memory throughput.  With this
approach, we were able to fully overlap the transpose with useful work
on the GPU. Therefore, the transpose is moved completely off the
critical path.

A drawback of this approach is that it puts additional stress on the
host CPU and memory bus. In a production setup, the host is also used
to receive the sample data from the network. 
Moreover, the CPU could
be used to perform additional tasks, such as pre- or
post-processing of the data. Possible steps could be filtering,
bandpass and phase corrections, etc.  It is unclear what would be more
efficient: using the host for the data transpose as described above, or
using it for additional processing steps. Further research is needed
to clarify thus further.

With the alternative memory layout, the performance of the kernel itself
increases from 314 to 357 gflops (38\% of the peak). 
For the version that includes the PCI-e I/O, the
performance increases from 274 to 300 gflops (32\% of the peak).
The internal memory bandwidth we achieve is 139~GB/s. This thus is
significantly higher than the 122~GB/s we got without the data transpose.
In Section~\ref{hardware-comparision}, we calculated that the
$\mathit{perf_{max,global}}$ for our hardware is 363 gflops. In
practice, we achieve 83\% of this limit due to the external I/O
problems.

\subsection{The Cell Broadband Engine (QS21 blade server)}


The basic \mbox{Cell/B.E.} programming is based on multi-threading:
the PPE spawns threads that execute asynchronously on SPEs.
The SPEs can
communicate with other SPEs and the PPE, using mechanisms like signals and mailboxes
for synchronization and small amounts of data, or DMA transfers for
larger data.  With the \mbox{Cell/B.E.} it is important to exploit all levels of parallelism.
Applications deal with task and data parallelism across multiple SPEs, vector parallelism
inside the SPEs, and double or triple-buffering for DMA
transfers~\cite{cell}.  The \mbox{Cell/B.E.} can be
programmed in C or C++, while using intrinsics to exploit vector
parallelism.

The large number of registers (128 times 4 floats) allows a big tile size of 
$4\times4$, leading to a lot of data reuse.
We exploit the vector parallelism of the \mbox{Cell/B.E.} by computing the four
polarization combinations in parallel.  We found that this performs
better than vectorizing over the integration time.  This is thanks to the \mbox{Cell/B.E.}'s
excellent support for shuffling data around in the vector registers.
\longversion{
This is useful for our application, since the different polarizations
and real and imaginary parts of the samples have to be multiplied and
added in several different ways.  
}
The shuffle instruction is executed
in the odd pipeline, while the arithmetic is executed in the even
pipeline, allowing them to overlap.

We identified a minor performance problem with the pipelines of the
\mbox{Cell/B.E.}  Regrettably, there is no (auto)increment instruction in the odd
pipeline.  Therefore, loop counters and address calculations have to
be performed on the critical path, in the even pipeline. In the time
it takes to increment a simple loop counter, four multiply-adds, or 8
flops could have been performed. To circumvent this, we performed loop
unrolling in our kernels. This solves the performance problem, but has
the unwanted side effect that it uses local store memory, which is
better used as data cache.

A distinctive property of the architecture is that cache transfers are
explicitly managed by the application, using DMA. This is unlike other 
architectures, where caches work transparently.
Although issuing explicit DMA commands complicates programming,
for our application this is not problematic.

By dividing the
integration time into smaller intervals, we can keep the sample data
for \emph{all stations} in the local store.  
We overlap communication with computation, by using multiple buffers.
For the sample data we use double buffering.
Thanks to the explicit cache,
the correlator implementation fetches each sample from main memory
\emph{only exactly once}. 

For the correlation output data, several approaches are possible.
In~\cite{ics}, we describe an implementation that loads and stores a strip of
tiles into the local store with one DMA operation.  Consider the
example of Figure~\ref{fig-correlation}, for instance, with a tile
size of $2\times3$.  In this case, we load three rows of correlations
(the height of a tile) at once, for instance the rows 0, 1, and 2.
Because of this, we have to load and store the correlations to main
memory several times, since the sub-results have to be accumulated.
Since the correlations are both read and written, we use triple
buffering in this case.

We also developed a version that keeps the correlation results in the
local store, and stores the correlation output data to host memory
only once. Double buffering is enough in this case. We can fit the
result in the local store if the number of receivers is not too large
(e.g., 64). For the LOFAR instrument, this is the case.  The achieved
performance of the two versions is identical. However, both have their
own advantages. The first version scales to larger numbers of
receivers, but uses more bandwidth between the host memory and the
SPEs. This bandwidth is available on the \mbox{Cell/B.E.}, so this
does not hurt performance.  The second version uses less bandwidth,
but is limited in the number of receivers.  However, the amount of
host memory bandwidth used is important, since we may want to run
additional operations on the host (the PPE), such as receiving data
from the network, and preprocessing it.


If an even larger number of receivers is used (e.g., 256), 
good performance can still be achieved on the \mbox{Cell/B.E.}.
This can be done by splitting the correlation triangle of
Figure~\ref{fig-correlation} in blocks (e.g., of size $32\times32$), which are
in turn divided into the tiles we already used in the previous
implementations.  On the SPEs, we now only load the samples inside the
block into the local store, and not the entire triangle.  A complication is that the X and
Y directions of the blocks no longer deal with the same samples in all
cases. For example, in the X direction, the samples could run from
0--31, while the Y-axis runs from 64--96. Therefore, the number of
samples that has to be loaded into the local store increases with a
factor of two. Nevertheless, the \mbox{Cell/B.E.} has the bandwidth
that is needed for this.

\longversion{
The performance we achieve on a single
\mbox{Cell/B.E.} chip is shown in Table~\ref{results-cell}.
}
Due to the high
memory bandwidth and the ability to reuse data, we achieve 188
gflops, including all memory I/O. This is 92\% of the peak
performance on one chip.  If we use both chips in the cell blade, the
performance drops only with a small amount, and we still achieve
91\% (374 gflops) of the peak performance.  Even though the memory
bandwidth per operation of the \mbox{Cell/B.E.} is eight times lower than
that of the BG/P, we still achieve excellent performance, thanks to
the high data reuse factor.

\section{Comparison and Evaluation}
\label{sec:perf-compare}

In this section, we compare the performance, power efficiency, and
programmablity of the different architectures. We also discuss a relatively
new development: OpenCL. We describe how we implemented the correlator using this language.

\subsection{Performance}
\begin{figure*}[t]
\begin{center}
\includegraphics[width=\columnwidth]{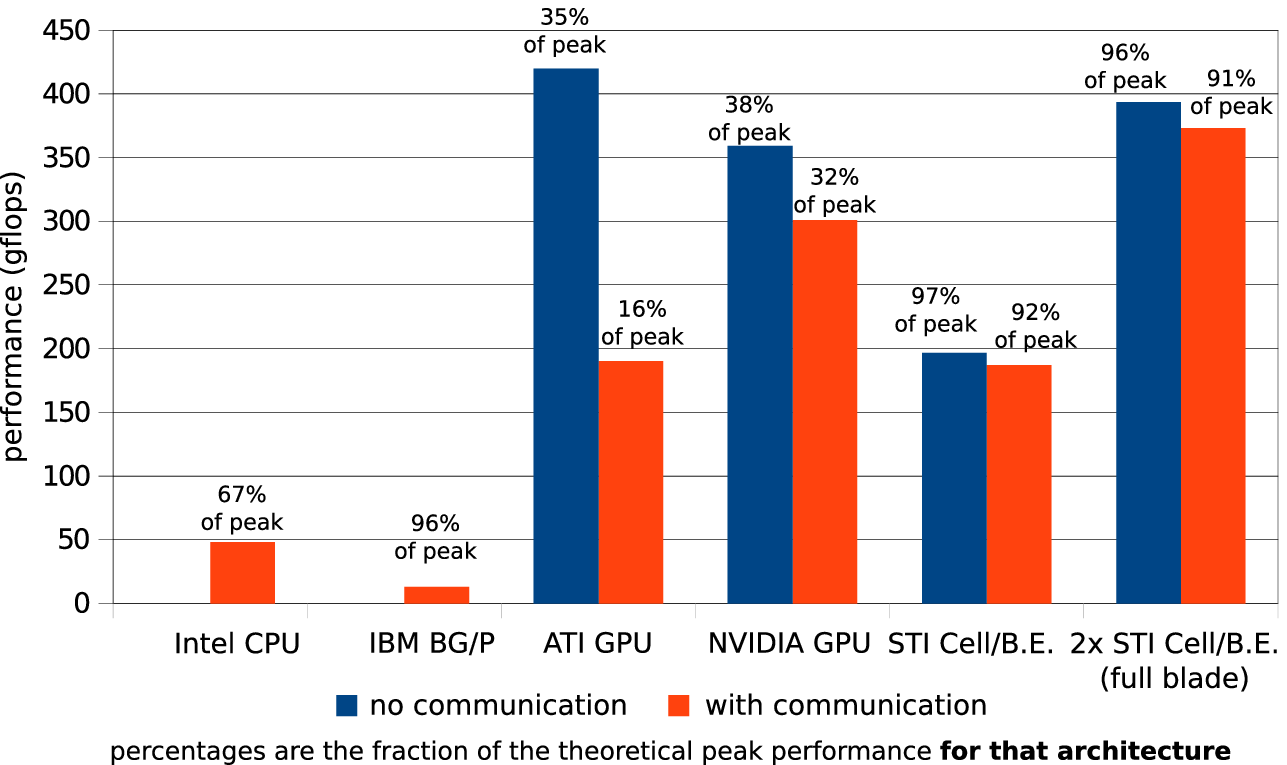}
\end{center}
\caption{Achieved performance on the different platforms.}
\label{performance-graph}
\end{figure*}

Figure~\ref{performance-graph} shows the performance on all
architectures we evaluated. The NVIDIA GPU achieves the highest
\emph{absolute} performance. Nevertheless, the GPU \emph{efficiencies}
are much lower than on the other platforms.  The \mbox{Cell/B.E.}
achieves the highest efficiency of all many-core architectures, close
to that of the BG/P. Although the theoretical peak performance of the
\mbox{Cell/B.E.} is 4.6 times lower than the NVIDIA chip, the absolute
performance is only 1.6 times lower.  If both chips in the QS21 blade
are used, the \mbox{Cell/B.E.} also has the highest absolute
performance. For the GPUs, it is possible to use more than one chip as
well.  This can be done in the form of multiple PCI-e cards, or with
two chips on a single card, as is done with the ATI 4870x2
device. However, we found that this does not help, since the
performance is already limited by the low PCI-e throughput, and the
chips have to share this resource.  The graph indeed shows that the
host-to-device I/O has a large impact on the GPU performance, even
when using one chip.  With the \mbox{Cell/B.E.}, the I/O (from main
memory to the Local Store) only has a very small impact.

\subsection{Power Efficiency}

\begin{table}[t]
\caption{Measured performance of the different many-core hardware platforms.}
\label{architecture-measurements}
{\small
\begin{tabular}{l|l|l|l|l|l}
                                          &  Intel       & IBM          &  ATI   &  NVIDIA       & STI Cell     \\
Architecture                              &  Core i7     & BG/P         &  4870  &  Tesla C1060  & (full blade) \\
\hline
measured gflops (including I/O)           & 48.0         & 13.1         & 190    &  300          & 374          \\ 
achieved efficiency                       & 67\%         & 96\%         & 16\%   &  32\%         & 91\%         \\
\hline
measured bandwidth (GB/s)                 & 18.6         & 6.6          & 81     &  139          & 49.5         \\
bandwidth efficiency                      & 73\%         & 48\%         & 70\%   &  136\%        & 192\%        \\
\hline
TDP chip (W)                              & 130          & 24           & 160    & 236           & 140          \\
theoretical gflops / Watt                 & 0.65         & 0.57         & 7.50   & 3.97          & 2.93         \\
achieved gflops/Watt (TDP)                & 0.37         & 0.54         & 1.19   & 1.27          & 2.67         \\
power efficiency compared to BG/P (TDP)   & 0.69 x       & 1.0 x        & 2.20 x & 2.35 x        & 4.9 x        \\
\hline
\textbf{measured power, full system} (W)  & 208          &  44          & 259    & 250           & 315          \\
\textbf{achieved gflops/Watt full system} & 0.23         & 0.30         & 0.73   & 1.20          & 1.18         \\
\textbf{power efficiency compared to BG/P}& 0.77 x       & 1.0 x        & 2.4 x  & 4.0 x         & 3.9 x        \\
\end{tabular}
}
\end{table}

In Table~\ref{architecture-measurements}, we present the power efficiency for
the different architectures.
The results show that the \mbox{Cell/B.E.} chip is
about five times more energy efficient than the BG/P.  This is not a
fair comparison, since the BG/P includes a lot of network hardware on
chip, while the other architectures do not offer this. Nevertheless,
it is clear that the \mbox{Cell/B.E.} is significantly more efficient.
A 45 nm version of the \mbox{Cell/B.E.} has been announced for early 2009. 
With this version, which has identical performance, but reduces the TDP to about 50W, 
the \mbox{Cell/B.E.} chip even is seven times more efficient than the BG/P.
The 65 nm version of the \mbox{Cell/B.E.} chip already is
about 2.3 times more energy efficient than the GPUs (based on TDP).
The fact that the three most energy efficient supercomputers on the
Green500 list\footnote{See http://www.green500.org.} are based on the
\mbox{Cell/B.E.} supports our findings. 
The Green500 list also specifies the achieved power efficiency for
entire supercomputers, i.e. including memory, chipsets, networking hardware, etc.
PowerXCell-based systems achieve 0.54 glops/W, while the Blue Gene/P
is less power efficient, and achieves 0.37 gflops/W. Systems based on
general-purpose CPUs only achieve 0.27 gflops/W.

The situation is different if we do not just look at the chips
themselves, but also take the full host system into account.  The
lower three lines in Table~\ref{architecture-measurements} show the
\emph{measured} power and efficiencies of the \emph{full systems},
while utilized.  For the GPUs, this means that we integrated them in
the Core~i7 host system that we also use for the general purpose CPU
measurements.  We measured the dissipated power with a VoltCraft-3000
power meter.  The results show that the full Cell blade uses
relatively much power in addition to the TDP of the two
\mbox{Cell/B.E.} chips.  This is partially caused by the XDR memory in
the system. As a result of this, the achieved performance per Watt is
lower than expected, but still 3.9 times higher than that of the
BG/P. For the NVIDIA GPU, however, the opposite is true.  In practice,
the device consumes much less power than specified by the TDP.
Therefore, the achieved efficiency of a full system with the GPU is
almost the same as that of the \mbox{Cell/B.E.}.  The ATI GPU is much
less efficient, due to the small fraction of the theoretical peak
performance that is reached in practice.

\subsection{Programmablity}

The performance gap between assembly and a high-level programming language 
is quite different for the different platforms. It also
depends on how much the compiler is helped by manually unrolling
loops, eliminating common sub-expressions, the use of register variables,
etc., up to a level that the C code becomes almost as low-level as assembly
code. The difference varies between only a few percent to a factor of 10. 

For the BG/P, the performance from compiled C++ code was by far not
sufficient.  The assembly version hides load and instruction
latencies, issues concurrent floating point, integer, and load/store
instructions, and uses the L2 prefetch buffers in the most optimal
way.  The resulting code is approximately 10 times faster than C++
code.  For both the Cell/B.E. and the Intel core~i7, we found that
high-level code in C or C++ in combination with the use of intrinsics
to manually describe the SIMD parallelism yields acceptable
performance compared to optimized assembly code.  Thus, the programmer
specifies which instructions have to be used, but can typically leave
the instruction scheduling and register allocation to the compiler.
On NVIDIA hardware, the high-level Cuda model delivers excellent
performance, as long as the programmer helps by using SIMD data types
for loads and stores, and separate local variables for values that
should be kept in registers. With ATI hardware, this is different.  We
found that the high-level Brook+ model does not achieve acceptable
performance compared to hand-written CAL code.  Manually written assembly 
is more than three times faster. Also, the Brook+ documentation is insufficient.

\begin{table}[t]
\caption{Strengths and weaknesses of the different platforms for data-intensive applications.}
\label{architecture-results-table}
{\small
\begin{tabular}{l|l|l|l|l}
Intel                        & IBM                              & ATI                          & NVIDIA                       & STI                           \\
      Core i7 920            &     Blue Gene/P                  &     4870                     &        Tesla C1060           &      Cell/B.E.                \\
\hline
+ well-known                 & + L2 prefetch unit               & + largest number             & + random                     & + random                      \\
                             & ~~~works well                    & ~~~of cores                  & ~~~write access              & ~~~write access               \\
                             & + high memory                    & + swizzling                  & + Cuda is                    & + shuffle                     \\
                             & ~~~bandwidth                     &                              & ~~~high-level                & ~~~capabilities               \\
                             &                                  &                              &                              & + explicit cache              \\
                             &                                  &                              &                              & + power efficiency            \\
                             &                                  &                              &                              &                               \\
                             &                                  &                              &                              &                               \\
\hline
- few registers              & - everything                     & - low PCI-e                  & - low PCI-e                  & - multiple         \\
- no fma                     & ~~~double precision              & ~~~bandwidth                 & ~~~bandwidth                 & ~~~parallelism levels                     \\
- limited                    & - expensive                      & - transfer slows             &                              & - no increment                \\
~~~shuffling                 &                                  & ~~~down kernel               &                              & ~~~in odd pipe                \\
                             &                                  & - no random                  &                              &                               \\
                             &                                  & ~~~write access              &                              &                               \\
                             &                                  & - CAL is low-level           &                              &                               \\
                             &                                  & - bad Brook+                 &                              &                               \\
                             &                                  & ~~~performance               &                              &                               \\
                             &                                  & - not well                   &                              &                               \\
                             &                                  & ~~~documented                &                              &                               \\
\end{tabular}
} 
\end{table}

In Table~\ref{architecture-results-table} we summarize the
architectural strengths and weaknesses that we identified.  Although
we focus on the correlator application in this paper, the
results are applicable to applications with low flop/byte ratios in
general.

\subsection{Implementing the Correlator with OpenCL}

An interesting recent development is the OpenCL programming
model~\cite{opencl}.  OpenCL is an open standard for parallel
programming of heterogeneous systems, developed by the Khronos
Group. Many important industry partners participate in the effort.
Many-core vendors (e.g., AMD, IBM, Intel, ATI, NVIDIA) have
pledged to support OpenCL to increase both portability and
programmability of their hardware. The main idea of OpenCL is that
a single language is used to program the many-core hardware of all
different vendors.
We have implemented the correlator
application in OpenCL, and tested it with the latest available OpenCL
implementations. Currently, AMD provides OpenCL support for multi-core \emph{CPUs} (not
GPUs) in the second beta release of their stream SDK version
2.0. NVIDIA released OpenCL support for their \emph{GPUs} in a
closed beta program. Since both implementations are still in the beta
stages, we do not discuss performance, but only programmability and portability.

OpenCL consists of two parts: a host API
(to interface with the C program that runs on the CPU), and a special
C-based language to develop kernels that run on the accelerators.  The
host interface is very low-level, and is similar to ATI's CAL, or
NVIDIA's driver API.  The Cuda model provided by NVIDIA has a much
higher abstraction level than OpenCL, and is significantly easier to
use. AMD provides an additional set of C++ wrappers around OpenCL's C
interface, but this set is not standardized. An important aspect of
OpenCL is that it uses \emph{run-time compilation} for the kernels. 

To demonstrate the level of abstraction of the host part of the OpenCL programming model,
we will describe the many steps that are necessary to run a kernel.  An OpenCL program has to
create a \emph{context} and a \emph{command queue}. Next, the kernel
program has to be loaded from disk (or constructed in memory), and a
program object has to be created. Then, the program has to be
compiled, and a kernel object has to be created. Subsequently, the
arguments to the kernel invocation have to be set, using a call per
parameter. Similar to CAL and Cuda, all buffers have to be explicitly
allocated and transferred from the host to the device. Finally,
a domain has to be specified, indicating the number of global and local threads to use.
Only after all
these steps, can the kernel be launched.  Also, if texture-cached
memory is used, as we do with the correlator on GPUs, the programmer
has to explicitly create 2D or 3D image objects, which also use a
different set of calls for the data transfers. Here, the graphics
origins of OpenCL clearly shine through.

The language that is used to write the kernels is based on C, with
additional extensions.  The level of abstraction is similar to Cuda,
and of higher level than CAL and the assembly that is needed to program
the Cell or CPUs efficiently. A large step forward is that the
language has a \emph{standard interface to perform vectorization}, and also
supports swizzling to shuffle data around inside vectors.  Special
annotations are used to declare data structures in the different
memory regions (global, constant, local, and private).  Finally,
special operations have to be used to load data from texture-cached
memory.

\begin{table}[t]
\caption{Strengths and weaknesses of OpenCL.}
\label{opencl-table}
{\small
\begin{tabular}{l|l}
Strengths & Weaknesses \\
\hline
+ high-level kernel language     & - low-level host API \\
+ vectorization and swizzling    & - only C binding standardized (no C++, Java, or Python) \\
+ portability                    & - performance portability not solved \\
+ runtime compilation            & \\
\end{tabular}
} 
\end{table}

We wrote several versions of the correlator in OpenCL, using different
tile sizes, and using normal and texture-cached memory.  A large
benefit of OpenCL is that the code indeed does compile and run without
any changes on both CPUs and GPUs. The only exception was a correlator version
that uses the texture cache: AMD's beta version does not support this yet.
We assume this problem will be solved for the final release.

However, the problem of \emph{performance} portability is not fully solved
by OpenCL yet.  First, we had to vectorize the correlator kernel
code. With NVIDIA hardware (and the Cuda model), this is not
necessary. Therefore, NVIDIA's runtime compiler removes the
vectorization.  On the CPU, the vectorization did increase
performance.  Second, the different platforms have very different
memory models. The programmer still has to deal with these differences
to reach the best performance. The texture cache of the GPUs and the local store of
the Cell/B.E. are examples of this. Moreover, we found that our
algorithmic changes are still necessary: the different platforms need
different tile sizes to perform optimally, due to the different numbers of registers they
have.  Finally, the different architectures require very different
numbers of global and local threads: only a few global threads for
CPUs, and many thousands, divided over global and local threads, for
the GPU. Table~\ref{opencl-table} summarizes the strengths and weaknesses of
OpenCL that we identified. 

 




\section{Related Work}
\label{related}

Intel's 80-core Terascale Processor~\cite{terascale} was the first
generally programmable microprocessor to break the teraflop barrier. It
has a good flop/Watt ratio, making it an interesting candidate for
future correlators.

Intel's Larrabee~\cite{larrabee} (to be released) is another promising
architecture.  Larrabee will be a hybrid between a GPU and a
multi-core CPU.  It will be compatible with the x86 architecture, but
will have 4-way simultaneous multi-threading, 512-bit wide vector
units, shuffle and multiply-add instructions, and special texturing
hardware. Larrabee will use in-order execution, and will have coherent
caches.  Unlike current GPUs, but similar to the \mbox{Cell/B.E.},
Larrabee will have a ring bus for communication between cores and for 
memory transactions.

Another interesting architecture to implement correlators are
FPGAs~\cite{fpga-correlator}. LOFAR's on-station correlators are also
implemented with FPGAs. Solutions with FPGAs combine good performance with 
flexibility. A disadvantage is that FPGAs are relatively
difficult to program efficiently.
Also, we want to run more than just the correlator on our hardware.
LOFAR is the first of a new generation of software telescopes, and how the
processing is done best is still the topic of research, both in
astronomy and computer science. We perform the initial processing steps on
FPGAs already, but find that this solution is not flexible enough for
the rest of the pipeline. For LOFAR, currently twelve different
processing pipelines are planned. For example, we would like to do the
calibration of the instrument and pulsar detection online on the same
hardware, before storing the data to disk. We even need to support
multiple different observations simultaneously. All these issues
together require enormous flexibility from the processing solution.
Therefore, we restrict us to many-cores, and leave application-specific
instructions and FPGAs as future work.
Once the processing pipelines are fully understood, future
instruments, such as the SKA, will likely use ASICs. 

Williams et al.~\cite{peri} describe an auto-tuning framework for
multi-cores.  The framework can automatically perform
different low-level optimizations to increase performance.  However,
GPUs are not considered in this framework. We performed all
optimizations manually, which is possible in our case, since the
algorithm is relatively straightforward. More important, we found that
in our case, algorithmic changes are required to achieve good performance.
Examples include the use of different tile sizes, and vectorizing
over the different polarizations instead of the inner time loop.

A software-managed cache is used on the \mbox{Cell/B.E.} processor. GPUs
typically have a small amount of shared memory that can be used in a
similar way~\cite{gpu-cache}.  An important difference is that in the
\mbox{Cell/B.E.} the memory is private for a thread, while with GPUs all
threads on a multiprocessor share the memory. The available
memory per thread is also much smaller. We applied
the technique described in~\cite{gpu-cache}, but found it did not
increase performance for our application.

Wayth et al.\ describe a GPU correlator for
the Murchison Widefield Array (MWA)~\cite{mwa-gpu-correlator}.
They optimize their code by tiling the correlator triangle in one dimension
(a technique described by Harris et.al.~\cite{gpu-correlator-harris}),
while tiling in two dimensions, as we described in this paper, is much more efficient.
For instance, a 2x2 tile requires the same amount of operations as a 1x4 tile,
but performs fewer memory operations~(see table~\ref{tile-size-table}).
For larger tiles, the arithmetic intensity of two-dimensional tiles is even
better.
Also, the MWA GPU version does not use the texture cache, but shared memory.
We found that this was significantly slower.
Their claim that their GPU implementation is 68~times faster than their CPU
implementation is highly biased, since their CPU implementation is not
optimized, single threaded, and does not use SSE.
As a result, our CPU version is 48 times faster than their CPU version,
while our GPU version is 4.2 times faster than their GPU version (even
though our data rates are four times as high due to our larger sample sizes).
Hence, their GPU implementation is only 1.4 times faster than an optimized
CPU implementation, not~68 times.

\section{Discussion}
\label{sec:discussion}


A key application characteristic of the correlator is that it is extremely
regular. This means that we know exactly which memory is referenced at
what time. In this paper, we explained that this property makes many 
optimizations possible.  We also implemented several other signal-processing algorithms we did not
discuss here, albeit not on all many-core architectures. Most
of our conclusions hold for all
(data-intensive) applications. However, this paper does not compare the ability of the
architectures to cope with \emph{unpredictable} memory access
patterns. We know, for example, that a particular radio-astronomy
imaging algorithm (W-projection) exhibits random memory access, and as a result
performs poorly on at least some of these architectures, and probably
all~\cite{gridding-08}. Also, the software-manged cache of the \mbox{Cell/B.E.}
is less effective here, since the programmer cannot predict the
accesses in advance. Fortunately, not all applications behave so
unpredictably.  In general, we advocate that the focus for
optimizations for many-core architectures should be on memory
bandwidth, access patterns, and efficient use of the caches, even at
the cost of increased synchronization and extra computation.

In this paper, we focus on the maximal performance that can be achieved
with a single many-core chip. An exiting result we present here is
that even extremely data-intensive applications, such as the correlator,
can perform well on many-core architectures, in particular on the
\mbox{Cell/B.E.}. These results allows us to move forward, and bring up the
question of scalability: can we scale the results to a full system
that processes all telescope data?  In this context, it is important
to emphasize that the correlator itself is trivially parallel, since
tens of thousands of frequency channels can be processed
independently.  However, in case of an FX correlator, a major data
exchange is necessary \emph{prior} to correlation: each input contains all
frequency channels of a single receiver, but the correlator requires a
single frequency channel of all receivers. We implemented this for the
LOFAR correlator on the 3-D torus of the Blue Gene/P, where we
exchange all data asynchronously. Although an efficient implementation
is complex, the time required for this exchange
is small compared to the time to correlate the data. Moreover, the
data rates grow linearly with the number of receivers, while the
compute time of the correlator algorithm grows quadratically. 
We also experimented on a PC cluster with a Myrinet switch, which was
able to handle the all-to-all exchange at the required data rates.  On the Blue Gene/P, we can
scale the application to more than 10.000 cores. For more information, we
refer to~\cite{spaa-06,ppopp2010}.

\section{Conclusions}
\label{conclusions}
Current and future telescopes have high computational and I/O demands.  Therefore,
we evaluated the performance of the extremely data-intensive
correlator algorithm on today's many-core architectures. This research
is an important pathfinder for future radio-astronomy instruments.
The algorithm is simple, we can therefore afford to optimize and
analyze the performance by hand, even if this requires assembly,
application-managed caches, etc. The performance of compiler-generated
code is thus not an issue: \emph{we truly compared the architectural
performance}.

Compared to the BG/P, many-core architectures have a significantly
lower memory bandwidth \emph{per operation}.  Minimizing the number of
memory loads per operation is of key importance.  We do this by
extensively optimizing the algorithm for each architecture.  This
includes making optimal use of caches and registers.  A high memory
bandwidth per flop is not strictly necessary, as long as the
architecture allows efficient data reuse.
This can be achieved through caches, local stores and registers.

Only two architectures perform well with our application.  The BG/P
supercomputer achieves high efficiencies thanks to the high memory
bandwidth per FLOP.  The \mbox{Cell/B.E.} also performs excellently, even
though its memory bandwidth per operation is eight times lower.  We
achieve this by exploiting the
application-managed cache and the large number of registers, 
optimally reusing all sample data. 
The \mbox{Cell/B.E.} is about five to seven times more energy efficient than the
BG/P, if we do not take the network hardware into account.

It is clear that application-level control of cache behavior (either
through explicit DMA or thread synchronization) has a substantial
performance benefit, and is of key importance for data intensive
high-performance computing.  The results also demonstrated that, for
data-intensive applications, the recent trend of increasing the number
of cores does not work if I/O is not scaled accordingly.

\begin{acknowledgements}
This work was performed in the context of the NWO STARE
AstroStream project.  We gratefully acknowledge NVIDIA, and in
particular Dr. David Luebke, for providing freely some of the GPU
cards used in this work. Finally, we thank Chris Broekema, Jan David
Mol, and Alexander van Amesfoort for their comments on an earlier
version of this paper.
\end{acknowledgements}


\bibliographystyle{abbrv}

\bibliography{correlator}

\end{document}